\documentclass[twocolumn]{aastex631}

\usepackage{multirow}
\usepackage{wasysym}
\usepackage{enumerate}
\usepackage{amssymb} %maths
\shorttitle{Earth-Moon Analogs with HWO}
\shortauthors{Limbach et al.}
\graphicspath{{./}{figures/}}

\begin{document}
\title{Exomoons \& Exorings with the Habitable Worlds Observatory I: On the Detection of Earth-Moon Analog Shadows \& Eclipses }

\correspondingauthor{Mary Anne Limbach}
\email{mlimbach@umich.edu}

\author[0000-0002-9521-9798]{Mary Anne Limbach}
\affiliation{Department of Astronomy, University of Michigan, Ann Arbor, MI 48109, USA}

\author[0000-0002-0746-1980]{Jacob Lustig-Yaeger}
\affiliation{Johns Hopkins APL, 11100 Johns Hopkins Rd, Laurel, MD 20723, USA}

\author[0000-0001-7246-5438]{Andrew Vanderburg}
\affiliation{Department of Physics and Kavli Institute for Astrophysics and Space Research, Massachusetts Institute of Technology, Cambridge, MA 02139, USA}

\author[0000-0003-0489-1528]{Johanna M.~Vos}
\affiliation{School of Physics, Trinity College Dublin, The University of Dublin, Dublin 2, Ireland}

\author[0000-0002-9831-0984]{Ren\'{e} Heller}
\affiliation{Max Planck Institute for Solar System Research, Justus-von-Liebig-Weg 3, 37077 G\"{o}ttingen, Germany}

\author[0000-0002-3196-414X]{Tyler D. Robinson}
\affiliation{Lunar \& Planetary Laboratory, University of Arizona, Tucson, AZ 85721 USA}

\begin{abstract}
The highest priority recommendation of the Astro2020 Decadal Survey for space-based astronomy was the construction of an observatory capable of characterizing habitable worlds. 
In this paper series we explore the detectability of and interference from exomoons and exorings serendipitously observed with the proposed Habitable Worlds Observatory (HWO) as it seeks to characterize exoplanets, starting in this manuscript with Earth-Moon analog mutual events.
Unlike transits, which only occur in systems viewed near edge-on, shadow (i.e., solar eclipse) and lunar eclipse mutual events occur in almost every star-planet-moon system. The cadence of these events can vary widely from $\sim$yearly to multiple events per day, as was the case in our younger Earth-Moon system. 
Leveraging previous space-based (EPOXI) lightcurves of a Moon transit and performance predictions from the LUVOIR-B concept, we derive the detectability of Moon analogs with HWO.  We determine that Earth-Moon analogs are detectable with observation of $\sim$2-20 mutual events for systems within 10\,pc, and larger moons should remain detectable out to 20\,pc.
We explore the extent to which exomoon mutual events can mimic planet features and weather. We find that HWO wavelength coverage in the near-IR, specifically in the 1.4\,$\mu$m water band where large moons can outshine their host planet, will aid in differentiating exomoon signals from exoplanet variability.
Finally, we predict that exomoons formed through collision processes akin to our Moon are more likely to be detected in younger systems, where shorter orbital periods and favorable geometry enhance the probability and frequency of mutual events.

\end{abstract}


%%%%%%%%%%%%%%%%%%%%%%%%%%%%%%%%%%%%%%%%%%%%%%%%%%%%%%%%%%%%%%%%%%%%%%
\section{Introduction} \label{sec:intro}
The discovery of over 5,000 exoplanets in recent decades has significantly expanded our understanding of the universe. These findings offer a glimpse into the diversity of planetary systems, many of which are strikingly different from our own Solar System planets. While we've gained knowledge about the types, frequencies, and potential habitability of these distant planets, there remains a significant gap in our knowledge of exoplanetary systems: the detection of moons and rings around exoplanets.
While a few potential signals hint at the presence of exomoons \citep{Ben_Jaffel_2014,Bennett_2014,Kenworthy_2015,Miyazaki_2018,2018AJ....155...36T,Oza_2019,Gebek_2020,Fox_2020,Lazzoni_2020,Limbach2021,Benisty_2021,2022NatAs...6..367K}, a definitive detection has eluded us. 

Our own Solar System serves as a compelling reference as to the diversity and prevalence of moons we may find in other planetary systems. Moons here outnumber planets by a wide margin, each with unique characteristics. From Europa's icy surface and subsurface ocean \citep{2000Sci...289.1340K}, Enceladus' out-gassing of water vapor \citep{2006Sci...311.1422H}, Triton's crystalline surface \citep{1997Icar..127..354Q}, Io's volcanism \citep{1979Sci...204..951S}, to Titan's intriguing atmosphere and methanological cycle \citep{2006P&SS...54.1177A,2007Natur.445...61S}, the diversity is astounding. These moons, formed through various processes such as disk collapses or gravitational interactions { (e.g., collision or capture)}, provide key insights into planetary formation and evolution \citep{2006Natur.441..834C,2006Natur.441..192A,2011Icar..214..113N,2012ApJ...753...60O,  Miguel_2016, 2018MNRAS.475.1347M, 2018MNRAS.480.4355C, 2020A&A...633A..93R}. We can thus expect to learn a lot more about the nature of exoplanets once it becomes possible to find natural satellites around them.

If our Solar System's rich tapestry of moons is indicative, it's reasonable to presume that many exoplanetary systems could be teeming with their own moons. Such exomoons, likely varied in composition and conditions, could present a broad spectrum of environments. Some might even reside within the habitable zones of their host planets, potentially possessing conditions conducive to the formation of life \citep{2010ApJ...712L.125K,2014OLEB...44..239L,HellerBarnes_2015,2018MNRAS.479.3477H}. Even if they are not habitable themselves, moons may be a key component to stabilizing life on habitable planets.
Earth's obliquity would vary chaotically if not for the stabilizing effect of the moon\footnote{ This works for the Earth
and Moon only because the gas-giants are well
separated from the terrestrial planets \citep{2000orem.book..513W}.} \citep{Laskar1993}, and therefore moons are predicted to play a substantial role in the long term climate stability and habitability of rocky exoplanets \citep{Williams1997, Spiegel2009, Armstrong2014, Meadows2018}. 
Grasping their demographics and characteristics could reshape our understanding of terrestrial bodies, their formation and the role of moons for habitability.

Unfortunately, the detection of exomoons analogous to the moons in the Solar System is not easy with existing observatories. JWST is capable of detecting transits of exomoons when planet/moon systems transit the host star. In a few select cases where wide-orbit Jovian-analogs transit their host star detection of Galilean moon analogs may be possible with JWST \citep{2023AJ....166..208H,2024jwst.prop.6491V}. Additionally, for close-in terrestrial worlds transiting small stars, Moon-analogs might be detectable with JWST \citep{2024jwst.prop.6193V}.
However, studying directly imaged exoplanets would enable a whole host of new exomoon detection methods, as previously discussed in the literature \citep{Cabrera_2007, 2009AsBio...9..269M, 2014ApJ...796L...1H, Agol_2015, Heller_2016,  2017MNRAS.470..416F, Vanderburg_2018, Lazzoni_2020}.

The Habitable Worlds Observatory (HWO), proposed in the National Academies’ ``Pathways to Discovery in Astronomy and Astrophysics for the 2020s" Decadal Report { \citep[][henceforth ``Astro2020'']{2021pdaa.book.....N}}, is a concept for a large space telescope operating in infrared, optical, and ultraviolet wavelengths. The mission's primary objective is to locate and analyze habitable planets outside our solar system, aiming to directly image at least 25 potentially habitable worlds. 

The advent of HWO is likely to enable the wide-spread detection of exomoons around mid-orbital separation (1-10\,AU) exoplanets for the first time. Where exomoon detection is currently out of our reach, for HWO their presence will manifest in a multitude of ways. Although HWO will spatially resolve exoplanets from their host star, a planet and its moons will generally remain unresolved. In many cases, moons are likely to have a notable or even dominant presence in the blended planet-moon spectral energy distribution (SED), potentially becoming a nuisance as we attempt to detect biosignatures on exoplanets \citep{2014PNAS..111.6871R}. Our Moon outshines the Earth in some infrared spectral bands \citep{2011ApJ...741...51R}. Even small moons can outshine their host planet \citep{2004AsBio...4..400W}. This was beautifully illustrated in JWST NIRCam imagery of the Solar System where Triton outshines Neptune\footnote{\url{https://www.nasa.gov/solar-system/new-webb-image-captures-clearest-view-of-neptunes-rings-in-decades/}} and where Europa is comparable in brightness to Jupiter\footnote{\url{https://blogs.nasa.gov/webb/2022/07/14/webb-images-of-jupiter-and-more-now-available-in-commissioning-data/}}. 
Exorings are also capable of significantly altering the SED of a planet \citep{2004A&A...420.1153A,2004ApJ...616.1193B,2005ApJ...618..973D,2022ApJS..263...15C}. In our own Solar System, Saturn's rings outshine the planet in the near infrared\footnote{\url{https://blogs.nasa.gov/webb/2023/06/30/saturns-rings-shine-in-webbs-observations-of-ringed-planet/}}.

Beyond just the desire to learn about exomoon populations, it's critical to detect and characterize exomoons in order to successfully study biosignatures and achieve the science goals of HWO. We cannot be certain about biosignature detections without understanding the contributions from the unresolved surroundings of an exoplanet, including exomoons and exorings. Understanding and accurately modeling these complex planetary environments is key to HWO's success.

Fortunately, HWO will be capable of detecting exomoons in many cases via extrasolar or lunar eclipses in the lightcurves of imaged planets. In this paper, we detail the framework governing mutual events, and apply it to our Moon-Earth-Sun system. Next, we derive the detectability of exomoons using mutual events with HWO, discussing the limitations that arise from planet. Finally, we discuss our findings and summarize the impact of Earth-Moon analogs in the context of HWO observations.

%%%%%%%%%%%%%%%%%%%%%%%%%%%%%%%%%%%%%%%%%%%%%%%%%%%%%%%%%%%%%%%

\section{Mutual Events of an Earth-Moon Analog}

\subsection{Framework}

\begin{figure}
\centering
\includegraphics[width=0.35\textwidth]{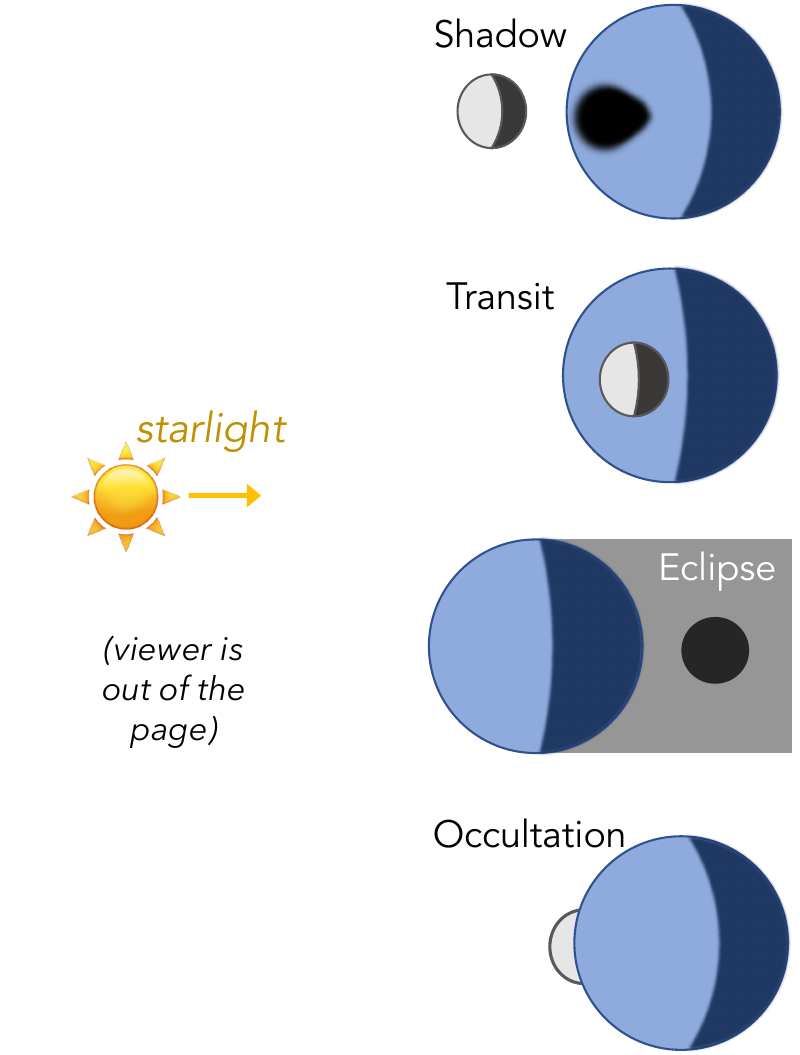}
\caption{An illustration of mutual events in a aligned moon-planet-star system viewed by an edge-on observer.}
\label{MutualEvents}
\end{figure}

When HWO monitors the reflected light lightcurves of an exoplanet hosting an exomoon, conjunctions between the moon-planet-star and moon-planet-viewer will produce mutual events. Conjunctions of the moon-planet-star arise in almost every system, regardless of geometry and viewing angle, and are depicted in Figure \ref{MutualEvents} (labeled shadows and eclipses). Adopting the terminology used for solar system moon mutual events and from \cite{2007A&A...464.1133C}, 
\begin{itemize}
    \item a {\it shadow} is when a moon passes between the star and planet, and
    \item an {\it eclipse} is when the planet passes between the star and moon.
\end{itemize} 
Two moon-planet-star conjunctions occur every moon orbit if a moon orbits near the planet-star ecliptic (as is the case for the Galilean satellites of Jupiter, which experience mutual events every $\sim$day; \citealt{2009A&G....50b..17G,2016A&A...591A..42S,2023Icar..39215348S,2023MNRAS.526.6145C}). If a moon is sufficiently inclined relative to the ecliptic, these events will typically occur multiple times per planet orbit.

If the orbital plane of an exomoon about the planet lies on the line-of-sight of the observer, then two additional mutual events, transits and occultations, will also occur every moon orbit -- these are the events which are currently routine for exoplanet detection and characterization.
\begin{itemize}
    \item An {\it occultation} occurs when the planet passes between the observer and the moon, and
    \item a {\it transit} occurs when the moon passes between the observer and the planet.
\end{itemize}
These events are also illustrated in Figure \ref{MutualEvents}.

\subsection{Signal Amplitude}

\subsubsection{Formalism}

Using the relative brightness of the moon and planet we can compute the depths of mutual events of an Earth-Moon system analog as seen by an edge-on viewer. During eclipse and occultations, a moon is always completely blocked by the planet (assuming a non-grazing geometry). The eclipse and occultation depth, $\delta_{occ}$, is given by
\begin{equation}\label{occEQ}
\delta_{occ} = \frac{F_{m}}{\left( F_{m} + F_{p} \right)},
\end{equation}
where $F_{m}$ is the flux of the moon and $F_{p}$ is the flux of the planet. Although the total brightness of the planet-moon system changes with the phase of the two bodies as they orbit their star, the eclipse and occultation depths remains the same under the assumption that the same fraction of the planet's and moon's disk are illuminated as seen by a distant observer (valid for $a_m<<a_p$). { We also note that this assumes that other higher-order effects are negligible. For instance, one could envision a scenario where the lunar phase function primarily exhibits a strong forward scattering peak, while the planet's behavior approximates Lambertian reflectance. Such conditions would result in a markedly distinct phase function for both the moon and the planet.}

The shadow and transit depths are a bit more complex as we must account for the flux from both objects as well as the phase of the planet as a moon only blocks a portion of the planet's disk. Assuming a uniformly illuminated disk, the fraction, $p$, of the planet's and moon's disk that is illuminated by the star is given by \cite {1979JRASC..73..233L} as:
\begin{equation}\label{EqP}
p = \frac{1}{2}\cos{\left( \frac{2\pi t}{T} \right)} + \frac{1}{2},
\end{equation}
where $T$ is the orbital period of the planet-moon system about the star in days and $t$ is the number of days since full phase (i.e., when the planet and moon are directly behind the star). A version of this formula that includes Lambertian scattering is given in \cite{2007A&A...464.1133C}, but for this manuscript we keep this simpler, albeit less realistic, formulation. The planet and moon reach quarter phase ($p = \frac{1}{2}$, half illumination) for $\frac{t}{T} = \frac{1}{4}, \frac{3}{4}$. 
The transit depth, $\delta_T$ is given by the fraction of the {\it illuminated} portion of the planet's disk that is blocked by the moon during transit, which is
\begin{equation}\label{TdepthEq}
\delta_T = \frac{R_{m}^{2}}{p R_{p}^2}\frac{F_{p}}{\left( F_{m} + F_{p} \right)},
\end{equation}
where $p$ is given in equation \ref{EqP}, $R_{m}$ is the radius of the moon and $R_{p}$ is the radius of the planet. Equation \ref{TdepthEq}, assumes that the full disk diameter of the moon is smaller than the planet's illuminated cross section at the equator. For small crescent sizes, this assumption is not valid, and a more complex equation is required. However, in practice, those cases are unlikely to be observable with HWO: near full and new phase the angular separation between the planet-moon system and the star will reside inside the coronagraph inner working angle.

The depth of a shadow cast on the planet as a moon passes between the star and planet is similar to the transit depth. Although the elongated shadow seen by an observer induces a slightly different change in flux, for the calculations in this manuscript, we will approximate these depths as equivalent.
More detailed equations describing the structure of mutual events are derived in previous literature \citep{1979JRASC..73..233L,2002JRASC..96...18F,2005JRASC..99...92F,2007A&A...464.1133C,2015IJAsB..14..191S} or for a more general formulation and modeling tools see \cite{2019MNRAS.483.3919V} and \cite{2022AJ....164....4L}, respectively.

\subsubsection{Spectral Dependence of the Signal Amplitude}\label{specDep}

Unlike the flux change of exoplanet transits of stars, which are roughly the same magnitude across spectral bands, the flux change due to mutual events can vary by orders of magnitude as a function of wavelength. To grasp the large range of signal amplitudes that can arise from mutual events, we investigate the Earth-Moon system as a function of phase and spectral band. However, we emphasize that the SEDs of terrestrial planets and moons in the Solar System are already extremely diverse prior to even considering terrestrial planets and moons beyond our Solar System. Thus the analysis achieved by narrowing our view to the Earth-Moon system alone is limited, but yet still informative.

To derive the spectral dependence of the signal amplitude, we use the Earth and Moon spectra measurements from the EPOXI mission \citep{2009ApJ...700..915C,2011AsBio..11..393R,2011AsBio..11..907L,2011ApJ...729..130C}. 
The EPOXI mission reused the Deep Impact flyby spacecraft for two key studies: the Extrasolar Planetary Observation and Characterization (EPOCh) and the Deep Impact eXtended Investigation (DIXI). These two components collectively formed the EPOXI mission. The Earth observations from EPOXI were conducted by observing the full Earth globe directly from the distant spaceborne platform situated in a heliocentric orbit along the equatorial plane, utilizing the same detectors each observation. Earth was monitored several times for the full 24-hour rotational cycle at three distinct epochs. This approach yielded time-resolved and time-averaged disk-integrated spectroscopy in the visible-to-near-infrared range and comprehensively charted the rotational lightcurve at various wavelengths.

\begin{figure}
\centering
\includegraphics[width=0.44\textwidth]{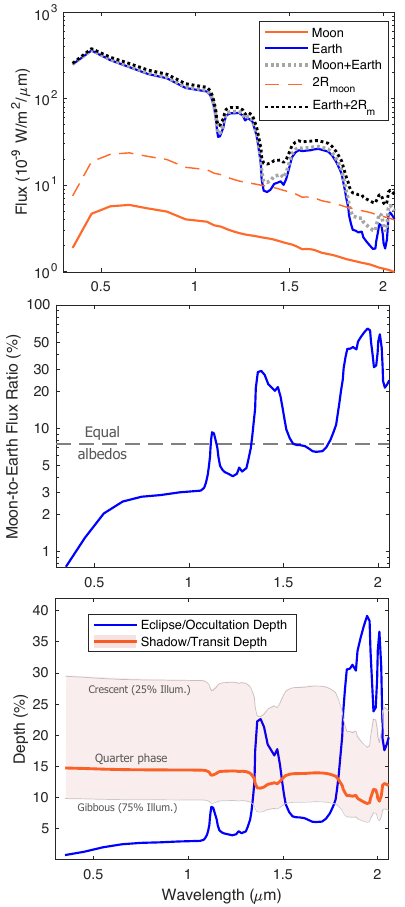}
\caption{{\bf Top:} Flux from the Earth (blue line), the Moon (red) and a moon with R = 2\,R$_{\rm Moon}$ (red dashed line) plus blended SEDs of the Earth and moons (dotted lines) as measured by the NASA EPOXI mission. {\bf Middle:} The Moon-to-Earth Flux ratio (blue line), which demonstrates the Moon is significantly fainter than the Earth in the UV/visible, but reaches 64\% of the apparent brightness of the Earth in the near infrared. For the case of equal albedoes the Moon would be 7.4\% the brightness of the Earth (gray dashed line). {\bf Bottom:} Depth of mutual events created by the Earth and Moon. The transit and shadow depth varies greatly depending on phase, but is relatively stable with wavelength. Occultation and eclipse events remain a constant depth with phase, but vary in depth by more than an order of magnitude with wavelength.}
\label{Spectra}
\end{figure}

Figure \ref{Spectra} shows the disk integrated flux of the Earth and Moon (top) taken from the EPOXI measurements, the ratio of the Moon-to-Earth flux (middle) and the depth of mutual events (bottom). The Moon's disk has 7.4\% the area of the Earth's disk, so in the case where they have equal albedoes (see dashed line on middle plot), the moon will produce a 14.8\% transit depth at quarter (or three-quarter) phase. From the middle plot, we see that the Moon is redder than the Earth. This results in eclipse depths much larger than 7.4\% at redder wavelengths. In fact, further in the infrared, in the 6.3\,$\mu m$ water band, the moon substantially outshines the Earth \citep{2011ApJ...741...51R}, producing eclipse depths near 90\%. However, we limit this study on wavelengths between $\lambda = 0.3-2$\,$\mu m$ as HWO is likely to operate in the NUV, optical and near infrared \citep{2019arXiv191206219T}. 

In the top panel of Figure \ref{Spectra}, we also include the SED of a moon twice the radius of our Moon (which would be about 10$\times$ the mass of the Moon or 0.1M$_\oplus$). While our Moon contributes to the blended Earth+Moon SED in the NIR, in the case of the larger moon, the Earth+2\,R$_{\rm Moon}$ NIR SED is dominated by the moon's flux. Although we defer exploring the impact of moons on our ability to accurately retrieve exoplanet spectra to a future manuscript, the blended SED in Figure \ref{Spectra} suggests the effects are non-negligible and can decrease the equivalent width of planetary atmospheric absorption bands.

The Earth-Moon transit depth for phases that correspond to disk illuminations between 25\%-75\% are plotted in the bottom panel of Figure \ref{Spectra} (light red shaded region). Quarter phase is illustrated by the red line. 
The transit and shadow depths vary widely depending on phase, ranging between 6-30\%, although if observed near quarter phase and in the visible (which should be typical for HWO observations), the transit depth for an Earth-Moon analog is about 15\%, or 2$\times$ $({R_{m}}/{R_{p}})^2$. 
Impressively, an occultation by the moon produces the largest change in flux reaching nearly 40\% in the NIR. However, as can be seen in the top panel of Figure \ref{Spectra}, the total system flux, $F_{m}+F_{p}$, is low where the occultation depth peaks and thus despite the large occultation depth, this may not necessarily be the highest signal-to-noise (SNR) event to measure (see section \ref{DetectLim}). 
Next, we will delve into the duration and frequency of these mutual events.

\subsection{Duration}

The duration of the eclipse and the occultation events {\it excluding} ingress and egress, and in a system where the moon-planet-star-observer is perfectly co-planar, is:
%\begin{equation} 
\begin{equation}
D_{occ} = \frac{T_{m} R_{p} } { \pi \, a_{m} }.
\end{equation} 
where $T_{m}$ is the orbital period of the moon about the planet and $a_{m}$ is the semi-major axis of the moon. We note that for eclipse events, this equation is valid for $a_{m}<<a_{p}$. If this is not the case, a small correction factor is needed (see \citealt{2007A&A...464.1133C}). For the Earth-Moon system, this gives 3.5 hours. The expression for the duration of eclipse and occultation events that includes ingress and egress are much more complex as the moon's phase must be included. At quarter (or three-quarter) phase, ingress and egress are about 30 minutes each in duration.

The duration of a transit or shadow event, again excluding ingress and egress, and in a system where the moon-planet-star-observer is co-planar,  is given by
\begin{equation} 
D_T = \frac { T_{m} \, R_{p} \,(1 + \cos \frac{2 \pi t}{T}) } { 2 \, \pi \, a_{m} }.
\end{equation} 
For non co-planar systems/viewers, the duration expressions will need to be modified. For the Earth-Moon system at quarter (or three-quarter) phase this gives 1.7 hours. In this case, ingress and egress will always be the same duration, lasting about an hour each, although this depends on inclination.

\subsection{Frequency}

Large moons orbiting terrestrial planets generally from through impact events. Simulations show that these moons initially form close to their parent planet and gradually move outward due to tidal interactions \citep{2004Icar..168..433C, 2012ApJ...760...83S, 2022ApJ...937L..40K}. Consequently, we anticipate shorter orbital periods at younger ages, leading to more frequent mutual events. This pattern is exactly what has been observed in the Earth-Moon system.

Previous simulations of the Earth-Moon forming giant impact suggest that our Moon formed at a distance of about $3.8\,R_\oplus$ from the Earth's center, close to its Roche radius at the time \citep{Canup2004}. At that distance, the orbital period of the system was about 10.5\,hr.

\begin{figure}[]
\centering
\includegraphics[width=0.5\textwidth]{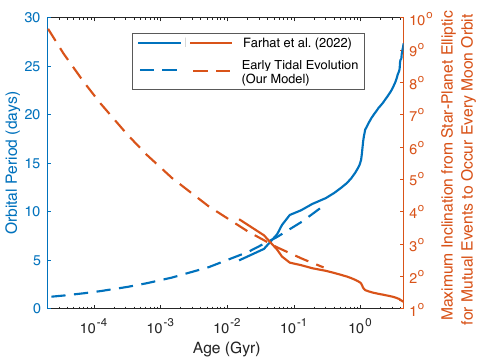}
\caption{The Moon's orbital period over time (blue dashed and { solid} lines) and the maximum inclination of the moon-planet relative to the star {\it or} a distant observer (red dashed and { solid} lines) for a mutual events to occur every moon orbit.  We used tidal evolution models to derive the orbital period at early times, and the Moon's orbital evolution at older ages was taken from \cite{2022A&A...665L...1F}.}
\label{MoonOrbitPer}
\end{figure}

The frequency of mutual events for the Earth-Moon system has changed drastically over time with the evolution of the Moon's orbital period and inclination. Currently, our Earth-Moon system undergoes a total of about 5 shadow (solar eclipse) or lunar eclipse events per year, but shortly after the moon formed, a shadow or eclipse event occurred every $\sim$6 hours. Similarly, if our Earth-Moon system was viewed near edge-on, transit and occultation events would currently occur every $\sim$two weeks, but right after formation these would have been once every $\sim$6 hours. The orbital period of our Moon and the maximum inclination allowed for mutual events to occur is illustrated in Figure \ref{MoonOrbitPer}.
 
Figure \ref{MoonOrbitPer} shows the evolution of the orbital period in the Earth-Moon system (left ordinate), with the results of our numerical tidal integration presented with a blue dashed line up to 300\,Myr (model is described in the following paragraph). Values from \citet{2022A&A...665L...1F} are used to complete the track up to the present. The orange lines illustrate the resulting (1) maximum orbital inclination of the system with respect to the line of sight of a distant observer for which transits and occultations events would occur, {\it or} (2) the maximum orbital inclination of the moon relative to the star-planet plane, including grazing events (right ordinate). In conclusion, this plot shows that for the first 300\,Myr the frequency of mutual events (whichever would be present for a given system) would be $\gtrsim$ one every ten days and occur for inclinations as high as $10^\circ$ initially and $\lesssim~2^\circ$ after 300\,Myr.

\subsubsection{Modeling the Tidal Evolution of our Moon}
For our numerical simulation, we apply the constant-phase-lag tidal model \citep{2008CeMDA.101..171F,2011A&A...528A..27H} to model the first 300\,Myr of the tidally driven orbital evolution of the Earth-Moon. Our parameterization of the system is identical to the one used by \citet{2021PalZ...95..563H}, except that here we use a tidal quality factor of $Q_\oplus~=~120$ since the resulting track of the orbital period evolution matches the one described by \citet{2022A&A...665L...1F}. The initial orbital period is 10.5\,hr and locked with the rotation of the Moon, whereas the rotation of the Earth is started at 2.2\,hr \citep{Canup2004,2021PalZ...95..563H}. The orbit is assumed to be circular with negligible spin-orbit misalignments. For the Moon, we chose a tidal quality factor of $Q_{\rm Moon}=40$ as a proxy for recent lunar ranging measurements \citep{2015JGRE..120..689W}. For both the early Earth and the early Moon we chose a second-degree tidal Love number of 0.3, thereby completing all the free parameters of the tidal model.

%%%%%%%%%%%%%%%%%%%%%%%%%%%%%%%%%%%%%%%%%%%%%%%%%%%%%%%%%%%%%%%%%%%%%%
%%%%%%%%%%%%%%%%%%%%%%%%%%%%%%%%%%%%%%%%%%%%%%%%%%%%%%%%%%%%%%%%%%%%%%
\section{Detectabilty of Moon-Analogs with HWO} \label{sec:detect}

HWO will monitor exoplanetary systems for days or weeks to detect and then characterize directly imaged exoplanets. The duration and cadence of these observations are amenable to the construction of reflected light lightcurves \citep{2001Natur.412..885F,2006AsBio...6..881T,2009ApJ...700..915C,2018AJ....156..301L,2024AJ....167...87S}. The constructed lightcurves will be comparable to the wide-orbit exoplanet and free-floating planet infrared lightcurves that have been observed with HST and Spitzer \citep[e.g.][]{Biller2018, Lew2020, Zhou2020, Vos2022} as well as ongoing time series observations with JWST \citep{2021jwst.prop.2327S,2023jwst.prop.2965B,2023jwst.prop.3496V,2023jwst.prop.3375W}, except in reflected light rather than emission and for much smaller planets, close to a host star. HWO will be incapable of spatially resolving exomoons and exorings that are orbiting directly imaged exoplanets. Consequently, the observed lightcurves will include flux contributions and variability originating from both the planet and its surroundings.

\begin{figure}
\centering
\includegraphics[width=0.46\textwidth]{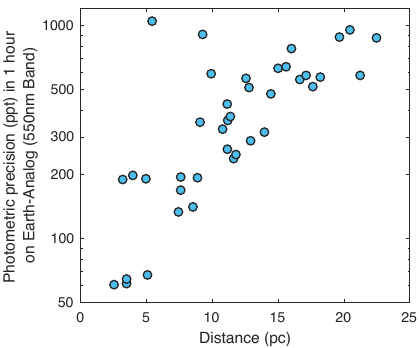}
\caption{The photometric precision (ppt) per hour on an Earth-analog in the habitable zone of 39 nearby systems observed with LUVOIR-B in the 550\,nm coronagraphic band. Photometric precisions are based on the calculations in \cite{2019arXiv191206219T}. }
\label{PhotPreEarth}
\end{figure}

\subsection{HWO Exomoon Detection Limits}\label{DetectLim}
In this section, we compute the number of mutual events that must be observed with HWO to detect (SNR = 5) a mutual event in a 550\,nm coronagraphic band which has a 20\% (110\,nm) spectral bandwidth.

\subsubsection{Methods}

\begin{figure*}
\centering
\includegraphics[width=0.96\textwidth]{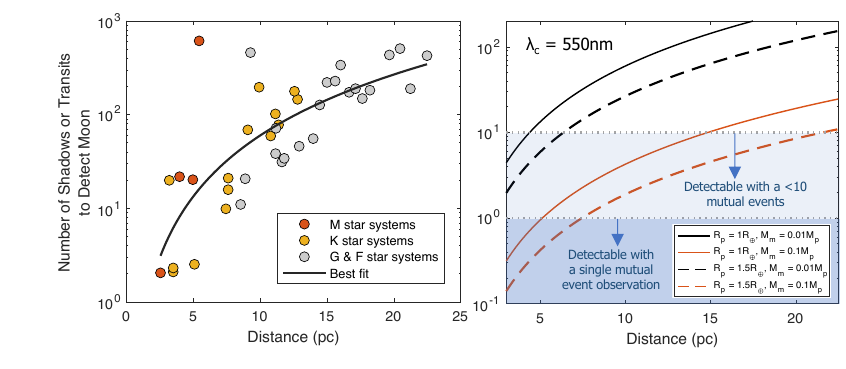}
\caption{{\bf Left:} Number of transits or shadow events required to detect (with a 5-$\sigma$ confidence level) a Moon-analog around an Earth-analog planet in 39 nearby star systems (circles) in the 550\,nm band, and the best fit to the simulations (black line). For the nearest systems, Moon-analogs are detectable with observation of just two shadow or transit events. The simulated systems leverage results from the LUVOIR-B concept in \cite{2019arXiv191206219T}. {\bf Right:} Number of transits or shadow events required to detect a moon with mass 0.01M$_{p}$ (black; our Moon is 0.01M$_\oplus$) or 0.1M$_{p}$ (red) around a planet that is 1.0\,R$_\oplus$ (solid lines) or 1.5\,R$_\oplus$ (dashed lines). In the most favorable cases, large moons of terrestrial worlds will remain detectable with $<$10 shadow or transit detections out to 20\,pc. Eclipse or occultation events are 10$\times$ harder to detect at this wavelength, but become much more amenable to detection at longer wavelengths (see Figure \ref{EclipseDetect}).}
\label{ShadowDetect}
\end{figure*}

First, to determine the photometric precision we can achieve on a typical terrestrial world, we leverage the LUVOIR-B sensitivity simulations shown in Figures 3-15 and 3-16 of The LUVOIR Mission Concept Study Final Report \citep{2019arXiv191206219T}. 
The simulations include coronagraphic speckles, both zodiacal and exozodiacal dust, telescope thermal emissions, photon noise, dark current, read noise, and clock-induced charge \citep{2016PASP..128b5003R,2019JOSS....4.1387L}. They are designed to calculate the necessary exposure time to achieve a specific photometric precision on an Earth-like planet in nearby star systems (within 25 pc). In the report, this is computed for 39 randomly chosen systems in specific spectral bands within the LUVOIR-B concept.  Here we leverage these computations to compute exomoon detectability, operating under the assumption that the performance of HWO will align closely with that of LUVOIR-B. This assumption is founded on the fact that HWO is planned to have a telescope aperture of comparable size to LUVOIR-B and shares the same primary scientific objective.

The photometric precision on an Earth-analog with one hour of observation in the LUVOIR-B 550\,nm coronagraphic band for the 39 systems is shown in Figure \ref{PhotPreEarth}. Using these photometric precision values, we calculate the number of transits or shadows that must be observed to reach an SNR = 5 on the transit or shadow event. The SNR of a single transit \citep{2023MNRAS.523.1182K} can be estimated as 
\begin{equation}\label{SNReq}
    SNR  \approx (\delta/\sigma_0)\sqrt{D},
\end{equation}
where $\delta$ is the transit depth, $\sigma_0$ is the photometric precision and $D$ is the duration of the transit. We assume the SNR will increase as $\sqrt{N_t}$, where $N_t$ is the number of transits observed. We note, however, that in the cases where the SNR of a single transit is low enough that folding multiple events is required for a confident detection, this estimate may be overly optimistic. This is because in this regime, it will be necessary to search for transits at many different trial periods, so the probability of false alarms will increase substantially due to multiple-hypothesis testing. In that case, we may need to achieve a higher SNR to confidently detect the planets than our nominal threshold of 5 \citep{Jenkins2002ApJ}.

We assume the system is observed when the planet is at quadrature and use equations \ref{EqP} and \ref{TdepthEq} to model the depth of transit or shadow event. For our computations, we use two planet sizes: R$_{p}$ = 1.0\,R$_\oplus$ and 1.5\,R$_\oplus$ and two moon masses: M$_{m}$ = 0.01\,M$_{p}$ and 0.1\,M$_{p}$. Here we assume that the 0.01\,M$_{p}$ moon is equivalent in mass and radius to our Moon, and in the M$_{m}$ = 0.1\,M$_{p}$ case, we take the moon radius to be 1.9$\times$ that of our Moon from the mass-radius scaling in \cite{2017ApJ...834...17C}.

Rather than assuming a a moon orbital period, we use a transit or shadow duration of 2\,hrs. Mutual event duration will vary depending on the system parameters, which will impact the number of mutual events required for detection. However, 2\,hrs is adopted as this is a typical duration for mutual events in our Earth-Moon system.

\subsubsection{Transit/Shadow Detection Results}

This results from our detection limit analysis are illustrated in Figure \ref{ShadowDetect}. Here we see that for a handful of the nearest systems, a shadow or eclipse is detected with observations of only two events. We provide a fit to the simulations (left panel) and then extrapolate this fit to the various sized planets and moons (right panel).

In the left panel, for systems out to $\sim$10\,pc, a Moon-analog is generally detectable with observations of $\lesssim$20 events. Collecting spectral measurements of exoplanets in the habitable zone is expected to require days to weeks of observation time. Thus, in young systems where Moon-analogs are on much shorter orbits, its quite plausible that observation duration will be sufficient to detect Moon-analogs out to 10\,pc. From the right panel, we see that for larger moons or planets, moon shadows and transits become much easier to detect. For some of these systems, detection remains plausible with $<10$ shadow or transit observations out to 20\,pc.

\subsubsection{Eclipse/Occultation Detection Results}

Figure \ref{ShadowDetect} only illustrates the detectability of shadow and transit events. At 550\,nm, the depth of eclipse and occultation events are an order of magnitude smaller than transits and shadows and will generally be undetectable except in the nearest systems. Referring back to Figure \ref{Spectra}, we recall that at 1.4\,$\mu m$ the eclipse and occultation events are generally larger in depth than transits and shadows. With this knowledge, we compute the number of eclipse and occultation events required to detect a Moon-analog using the same process as before, but now 1.4\,$\mu m$ using equations \ref{occEQ} and \ref{SNReq}.

\begin{figure}[b]
\centering
\includegraphics[width=0.46\textwidth]{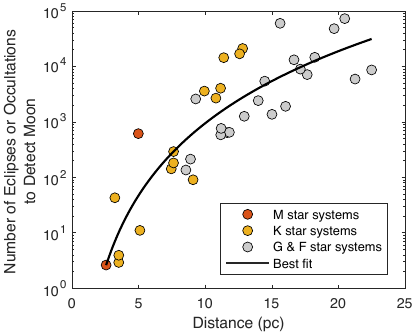}
\caption{Number of eclipse or occultation events required to detect (with a 5-$\sigma$ confidence level) a Moon-analog around an Earth-analog planet in 37 nearby star systems (circles) in the 1.4\,$\mu m$ spectral band, and the best fit to the simulations (black line). For the nearest systems, Moon-analogs are detectable with observation of just a few eclipse or occultation events. In this spectral band, eclipse or occultation events can be detected with half the number of shadow or transit events that would be required for detection. The simulated systems leverage results from the LUVIOR-B concept in \cite{2019arXiv191206219T}. }
\label{EclipseDetect}
\end{figure}

The results of this calculation are plotted in Figure \ref{EclipseDetect}. Impressively, we see that in the nearest systems, HWO still only requires observation of a few eclipses or occultations to detect a Moon-analog. However, in this case the number of events required for detection increases more steeply with distance to the system. This is likely due to $\lambda$/D increasing significantly in the near IR, which limits performance at these wavelengths, and thus NIR exomoon occultations/eclipse detections limits will likely be notably better for planet-moon systems orbiting beyond the habitable zone. In this 1.4\,$\mu m$ spectral band eclipses and occultations can be detected with half the number of events that would be required for shadow and transit events of an Earth-Moon analog.

\subsection{Detection Limits in the Presence of Exoplanet Variability}

\begin{figure}
\centering
\includegraphics[width=0.35\textwidth]{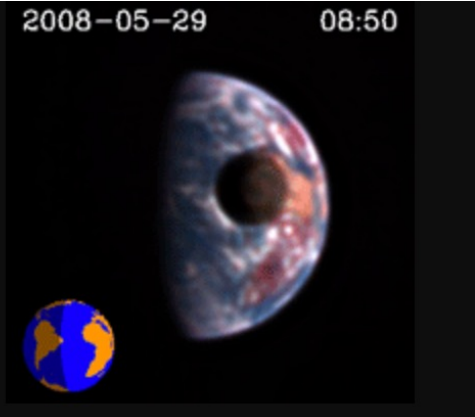}
\caption{Still frame of the Moon transit during NASA EPOXI mission observations of Earth. 
 Video link: \url{https://epoxi.astro.umd.edu/3gallery/vid_Earth-Moon.shtml}; Video Credit: Don J. Lindler, Sigma Space Corporation and NASA/JPL-Caltech/GSFC/UMD.}
\label{EPOXIstillFrame}
\end{figure}

In the previous section we established the detection limits assuming that the detection of mutual event signals will be limited by the observatory's sensitivity. However, terrestrial planets analogous to Earth will vary in flux due to landmasses and oceans rotating in and out of view \citep{2001Natur.412..885F,2006dies.conf..153G,2019arXiv191206219T}. There will also be variations due weather changes and cloud variation across the planet. In the presence of a highly rotationally variable exoplanet, precision may be limited by planet variability.
Therefore, we now explore: {\it What is the SNR, as a function of spectral band, required to detect a Moon-analog transiting an Earth-like planet in the presence of planet variability?}

\subsubsection{Methods}

Fortunately, a prime dataset is available to do just this. In 2008 the NASA EPOXI mission 
captured observations of the Earth during a Moon transit. Shown in Figure \ref{EPOXIstillFrame} is a still frame from the EPOXI observations of the Moon transiting the Earth.

\begin{figure}[b]
\centering
\includegraphics[width=0.48\textwidth]{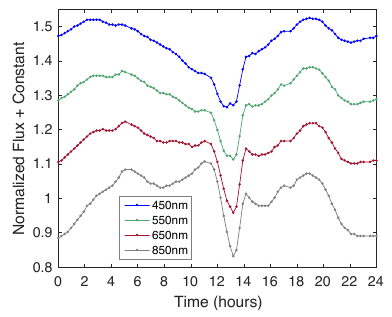}
\caption{Measured lightcurves of the Moon transiting in front of Earth as observed by the NASA EPOXI mission.}
\label{lightcurves}
\end{figure}

\begin{figure*}
\centering
\includegraphics[width=1\textwidth]{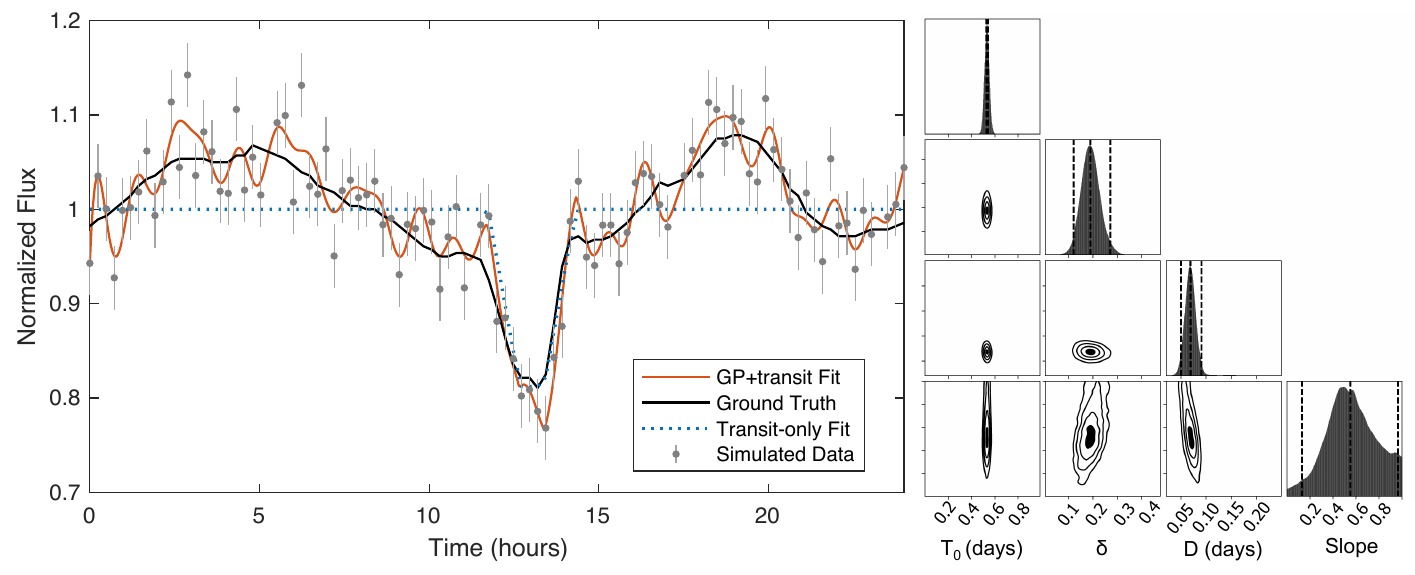}
\caption{{\bf Left:} An example of simulated data (gray errorbars) based on the EPOXI lightcurve measurement (black line) at 550\,nm with a simulated photometric precision of 4\% per 15 minutes (1.5\% precision for the full transit duration). The GP+transit fit (red line) and transit only fit (blue dotted line; from the GP+transit fit) is also plotted. {\bf Right:} Corner plot from GP+transit fit showing the mid-time of transit (T$_0$), the transit depth ($\delta$), the transit duration ($D$) and the slope of ingress/egress.}
\label{simLCandFit}
\end{figure*}

In \cite{2011AsBio..11..907L}, the unresolved Earth+Moon photometry was extracted from the data to produce lightcurves in multiple spectral bands. The resulting EPOXI lightcurves are shown in Figure \ref{lightcurves}. From these lightcurves, it is clear that both the Earth's rotational variability as well as the Moon transit contribute to significantly to observed amplitude variation.

Leveraging the EPOXI data, we simulate lightcurves with various levels of additional photon noise. 
We produce 400 simulated lightcurves varying the level of photon noise and the spectral band. We use 15\,min, 1\,$\sigma$ noise levels of  2\%, 4\%, 6\%, 8\% and 10\%. Although the HWO will typically not reach these simulated photometric precisions with a single transit observation as shown in the previous section, this analysis will still enable the determination of the exoplanet variability and observatory noise limited regimes. 

In the EPOXI lightcurves, the transit duration lasts 1.7\,hrs. Thus, the total photometric precision achieved on the transit is calculated by simply dividing our 15\,min photometric precision by $\sqrt{N_p}$, where $N_p$ is the number of 15\,min data points collected during the duration of the transit.

Using these simulated lightcurves, we can determine the regimes in which we are limited by photon-noise verses planet variability. To do this we search for transits in the simulated lightcurves.
This is accomplished by using a Gaussian process (GP) model to capture the rotational variability of the terrestrial planet and a trapezoidal transit model to fit the lunar transit. The transit model uses four parameters: mid-transit time, transit depth, transit duration and the impact parameter. The GP model is the same as that described in \cite{Limbach2021}. 

We then use \texttt{Dynesty}, a nested sampling code \citep{2020MNRAS.493.3132S}, to fit GP and transit parameters. We classify a Moon transit as detected if the mid-transit time is within 50\,min of the actual mid-transit time (so that the mid-transit time is required to be within the known duration of the transit). To determine the ground-truth mid-transit time, we use the GP+transit code fit from the EPOXI lightcurve prior to adding additional photon noise.

\subsubsection{Results}

\begin{figure}[t]
\centering
\includegraphics[width=0.5\textwidth]{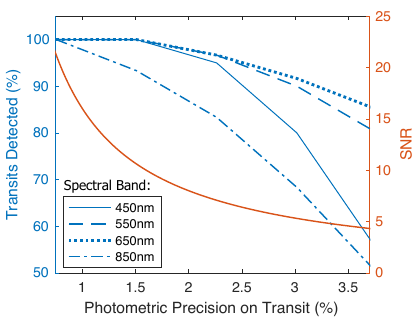}
\caption{{\bf Left Axis }{({\color{blue} blue} lines)}: Percentage of detected Moon transits as a function of photometric precision in the simulated lightcurves using the EPOXI data. Transits are more difficult to detect at 450\,nm and 850\,nm due to the structure of the Earth's variability in those spectral bands. {\bf Right Axis }({\color{purple} red} line): Theoretical SNR, from equation \ref{SNReq}, in the photon noise limit.}
\label{RetrivalRatePlt}
\end{figure}

An example of a simulated lightcurve at $\lambda$ = 550\,nm and the corner plot for the transit parameters retrieved from the nested sampling analysis are shown in Figure \ref{simLCandFit}. In this example, where a precision of 1.5\% is achieved for the full transit duration (4\% precision/15\,min) the transit is well constrained, with a fitted transit depth of 17.3$^{+2.3}_{-2.1}$\% compared to the actual transit depth of 14.3\%. At higher noise levels, the fits become less constrained or fails to find the transit completely. Figure \ref{RetrivalRatePlt} shows the percentage of detected transits (blue lines) as a function of the photometric precision achieved during the full transit duration for all four spectral bands.

The red line also plotted on Figure \ref{RetrivalRatePlt} is the theoretical SNR we would expect in the presence of no planet variability (photon-noise limited SNR). For the Moon, a photometric precision on the transit of 1\% corresponds to a SNR = 16, as defined by equation \ref{SNReq}. At this precision, nearly all transits in every spectral band are detected. Conversely, with a SNR = 5 (eq. \ref{SNReq}), which corresponds to a photometric precision of 3\% for the moon, 90\% of transits are still detectable at 550\,nm and 650\,nm, but significantly less are detected in the other two spectral bands.

Notably, at 550\,nm and 650\,nm, transits are more detectable, despite identical noise levels across all spectral bands in our simulations. As Figure \ref{lightcurves} illustrates, the planet's variability in the  lightcurve shows greater fluctuations at 450\,nm and 850\,nm. This is further evidenced by the more stable pre- and post-moon transit flux baselines at 550\,nm and 650\,nm, in contrast to the varying baselines at 450\,nm and 850\,nm. 

\begin{figure}[]
\centering
\includegraphics[width=0.36\textwidth]{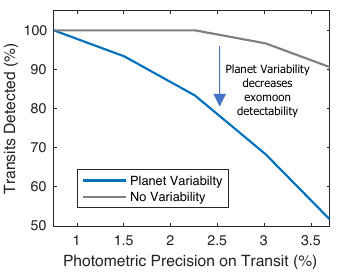}
\caption{Detectabilty of exomoon transits in the 850\,nm lightcurve with Earth's variability included in the lightcurve (blue line; same as the dash-dotted blue line in figure \ref{RetrivalRatePlt}) and with Earth's variability removed (gray line). In the case with variability removed, the moon remains detectable even as the photometric precision worsens (goes up). For our earth-moon system and at 850\,nm, this demonstrates that the detectabilty of our moon is limited primarily by host variability rather than photon noise. In other spectral bands or for larger moons that produce much larger transits than planet variability amplitudes, photon noise will be the primarily limitation in moon detection.}
\label{NoVar}
\end{figure}

To confirm that planetary variability hinders transit detection, we also simulated lightcurves at 850\,nm with removed host variability, normalizing the relative flux prior and post transit ingress and egress, respectively, to one. Analyzing these modified lightcurves with our GP+transit code, we observed a significant increase in transit detection rates (over 90\% success for photometric precisions $>3\%$ per transit) as shown in Figure \ref{NoVar}. This contrasts with only a 50\% detection rate under original host variability conditions. Remarkably, the 850\,nm lightcurve without host variability outperforms all bands with host variability, indicating that for a true Earth-Moon analog, the HWO's transit detection efficiency is more influenced by host variability than by photon noise. We conclude that while a SNR=5 on a mutual event is often sufficient for detection, in the presence of substantial planet variability, higher SNRs will be required for mutual event detection.

For terrestrial worlds with more or less uniform surfaces/atmospheres (e.g., Venus, Mars, Mercury, a pure water world with no clouds, etc.) variability will not impede exomoon detection. 
For larger moons, where the transit depths are significantly larger than the amplitude variations from the planet, the exomoon detectability will be significantly less impacted. Conversely, the detectability of moons smaller than our own Moon are likely to be even more severely limited by the planet's amplitude variations.

However, there may be techniques to improve moon detection. The timescale of a transit will generally be different than the compared rotation period of the planet. Presumably, most planets that rotate slow enough that they will appear to have relatively consistent flux levels over the duration of the moon transit.
Additionally, if lightcurves include spectrophotometric information rather than just a single broadband flux, it may be possible to use the spectral information to disentangle achromatic transits from chromatic host variability \citep{Limbach2021}.

Finally we note that the EPOXI lightcurves only include a single rotation of the Earth. It is likely that observations of several planet rotations will increase our ability to differentiate between moon transits and planet rotational variability, especially in the case where planet variability is driven by the distribution of landmasses and ocean rather than weather (clouds) and thus constant with each rotation. In this regard, moon detection may be less sensitive to planet variability than the calculations in this section would suggest.

%%%%%%%%%%%%%%%%%%%%%%%%%%%%%%%%%%%%%%%%%%%%%%

\section{Discussion} \label{sec:discuss}

\subsection{Can Moons Mimic Weather?}
 
Exomoons are likely to be a source of confusion when observing exoplanets with HWO. For example, consider a barren planet with a moon on a several day orbital period. If observing in the visible, a moon's transits and/or shadows could be mistaken for planet's rotation period: a feature (such as a land mass) on the planet's surface rotating in and out of view. Differentiating these two scenarios is possible, but would require additional observations. The most straightforward way to confirm a moon would be with the detection of occultation and eclipse events in the near-IR, which would shift in cadence by a factor of two relative to the eclipses and shadows detected in the visible, whereas a feature on the planet's surface would remain strictly periodic. 

Longer duration observations and high cadence observations would also aid in differentiation between the two scenarios as well. We expect variability (due to seasons and changes in the weather) and transit depth/duration to evolve (see Figure \ref{Spectra}) with lunar phase. Although these features will change in different ways, understanding or predicting that change may be difficult if the planet's surface and weather is complex and the lightcurves are low signal-to-noise.
Further modeling beyond the scope of this manuscript is needed to fully understand the extent to which this is will be a complicating factor.

From Figures \ref{EPOXIstillFrame} and \ref{lightcurves}, we can see that the moon first transits the ocean and then passes in front of a landmass (Africa). Because the oceans are significantly bluer than the land, the transit duration appears to be longer at bluer wavelengths. In this way, one could imagine that it may be possible to map the surface of the planet with a Moon transit on the occasion where sufficiently high SNR measurements are available in the same way we currently map starspots with transiting exoplanets \citep{2013MNRAS.428.3671T,2017A&A...597A..94C,2019A&A...630A.122A} or exoplanet daysides with eclipse mapping \citep{2007ApJ...664.1199R,2007Natur.447..183K,2012A&A...548A.128D,2019AJ....158..166B}.

\subsection{HWO's Sensitivity to Exomoons Orbiting Non-HZ Terrestrial Exoplanets}

Herein we have only computed the detectabilty of exomoons around terrestrial planets in the habitable zone (of which HWO is estimated to detect $\sim$25). However, HWO is likely to observe the lightcurves of many more terrestrial worlds including planets in multi-planet systems that are simultaneously observed while characterizing habitable zone terrestrial planets, as well as shorter observations obtained of every HWO target system during the search for systems which host habitable zone planets.  Based on the expected detection yield for LUVOIR-B (right panel, Figure 3-12 of the LUVOIR final report) is 68 rocky planets (radii of 0.5-1\,R$_\oplus$) and 127 super-Earths (radii of 1-1.75\,R$_\oplus$) \citep{2019arXiv191206219T}, there is likely to be a large number of terrestrial worlds and giant planets which can be searched for exomoons. 
{ A larger range of Solar System ground-truth observations \citep[e.g.,][]{2021DPS....5341208W} of planet-moon mutual events would aid in laying the ground work for understanding moon detectability within HWO lightcurves and would be beneficial for mission preparation}.

\subsection{Exomoon Science Questions that will be Addressed by HWO}

HWO will be capable of detecting and characterizing a large range of moons. It is critical that we determine now the tools and observatory requirements that are needed to disentangle moon and planet (and ring) signatures so that the HWO primary mission is not thwarted by the presence of exomoons. Not only will this ensure that HWO is capable of accomplishing its primary mission (the detection of biosignatures), it will also enable us to understand the breadth of exomoon science that HWO is capable of accomplishing.

The HWO will have the capability to answer critical questions related to exomoon science, including:
\begin{itemize}{\it
    \item What are the demographics of moons around giant planets and terrestrial planets at 1-10\,AU, and how does this population of exomoons compare to those within our Solar System?
    \item What is the occurrence rate of habitable-zone exomoons?
    \item  Does the presence of moons correlate with atmospheric characteristics, climate states, and habitability?
    \item  How will the presence (or lack) of exomoon detections inform our understanding of exoplanet formation via pebble accretion versus collisions? {\rm If exoplanet form via collisions, we expect a high occurrence of fractionally large moons orbiting terrestrial planets.}
    \item Are Moon-analogs critical to the formation of life on Earth-like planets?
    \item Can HWO detect biosignatures on exomoons?}
\end{itemize}   

These questions not only highlight the capabilities of HWO in exomoon research but also illustrate the broader implications of such discoveries for our understanding of planetary systems and the conditions conducive to habitability and life.

\section{Conclusion}

In this manuscript, we demonstrated that all four types of mutual events that can arise in an Earth-Moon analog system will be detectable with HWO under some conditions. 
We summarize the major findings of this manuscript as follows:
\begin{itemize}
    \item We conclude that a baseline HWO architecture could be capable of detecting exomoons analogous to our own Moon around terrestrial planets within 10\,pc with detection of $\sim$2-20 mutual events\footnote{The technical specifications for HWO are under development. In this manuscript we assume performance on-par with LUVOIR-B as reported in \cite{2019arXiv191206219T}. If HWO's specifications diverge significantly from LUVOIR-B, the detectability of exomoons may also change significantly.}. Exomoons that are larger and/or those that orbit larger planets may be detectable via mutual events out to 20\,pc, inclusive of most HWO target systems. Shadow and Eclipse events occur in almost all moon-planet-star systems with frequencies from hours to years, depending on their geometry, while transits and occultations only occur in systems viewed nearly edge-on.
    \item Moons that form via collisions will initially have short orbital periods leading to more probable and frequent mutual events. Hence, we predict exomoons around terrestrial worlds are more likely to be detected in young systems.
    \item Exomoon mutual events may mimic host variability and visa versa. HWO wavelength coverage in the near-IR, specifically in the 1.4\,$\mu$m water band where large moons can outshine their host planet, and shorter exposure cadence, ideally $\sim$15\,min, will help differentiate exomoon signals from planet rotational variability. Further, the NIR coverage would would be enhanced by tighter coronagraphic inner working angles.
    \item Exomoons (and exorings) have the potential to greatly alter or even dominate portions of the planet+moon SED, as well as produce significant signatures in planet-moon lightcurves. To ensure HWO is designed with the specifications needed to achieve its primary mission (the ability to detect biosignatures on habitable worlds) it is critical that we include moons and rings in our modeling. 
\end{itemize}

There are currently no constraints beyond our own Solar System on the occurrence rate of moons orbiting terrestrial planets at separations where we expect moons to be common \citep[e.g., $\gtrsim$1\,AU;][]{2002ApJ...575.1087B,2010ApJ...719L.145N,2018AJ....155...36T,2020MNRAS.499.1023I,2022MNRAS.513.5290D}. However, the results of numerical simulations suggests that fractionally-large exomoons may be common around terrestrial planets \citep{1997Natur.389..353I,2022NatCo..13..568N}. HWO will allow us to test this hypothesis by constraining the occurrence rate of this population of exomoons. The remarkable constraints that HWO can impose on the moons of terrestrial exoplanets are poised to revolutionize our present knowledge of exomoons.

%\hfill \break
\section*{Acknowledgement}
We thank Fred Adams and Bruce Macintosh for helpful conversations while preparing this manuscript. This research has made use of the NASA Exoplanet Archive, which is operated by the California Institute of Technology, under contract with the National Aeronautics and Space Administration under the Exoplanet Exploration Program. 
JLY and MAL were funded by 
the CHAMPs (Consortium on Habitability and Atmospheres of M-dwarf Planets) team, supported by the National Aeronautics and Space Administration (NASA) under Grant No. 80NSSC23K1399 issued through the Interdisciplinary Consortia for Astrobiology Research (ICAR) program. R.~H. acknowledges support from the German Aerospace Agency (Deutsches Zentrum f\"ur Luft- und Raumfahrt) under PLATO Data Center grant 50OO1501. J. M. V. acknowledges support from a Royal Society - Science Foundation Ireland University Research Fellowship (URF$\backslash$1$\backslash$221932).

\facilities{{\it The Habitable Worlds Observatory}}

\software{{\tt astro.py} (\url{https://github.com/astropy/astropy}), {\tt numpy.py} \citep{5725236}, {\tt dynesty.py}  \citep{2020MNRAS.493.3132S}, {\tt corner.py} \citep{corner} and {\tt ChatGPT} was utilized to improve wording at the sentence level; Last accessed February 2024, OpenAI (\url{chat.openai.com/chat}).}

\bibliography{main}{}
\bibliographystyle{aasjournal}
\end{document}